\documentclass[12pt]{article}
\pdfoutput=1
\pdfoutput=1
\usepackage[DIV13]{typearea}
\usepackage{indentfirst}
\usepackage{amsmath}
\usepackage{mathrsfs}
\usepackage{amssymb}
\usepackage{bbm}
\usepackage{graphicx}
\usepackage{braket}
\usepackage{slashed}
\usepackage{multicol}
\usepackage[usenames,dvipsnames]{color}
\usepackage{cite}
\usepackage{cancel}
\usepackage[normalem]{ulem}
\usepackage{array}
\usepackage{multirow}
\usepackage{wasysym}
\RequirePackage[colorlinks=true,urlcolor=blue,anchorcolor=blue,citecolor=blue,filecolor=blue,
linkcolor=blue,menucolor=blue,linktocpage=true,pdfproducer=medialab]{hyperref}
\newcommand{\email}[1]{\href{mailto:#1}{\tt #1}}

\numberwithin{equation}{section}
\newcommand{\LL}{\mathcal{L}}

\def\be{\begin{equation}}
\def\ee{\end{equation}}
\def\bc{\begin{center}}
\def\ec{\end{center}}
\def\bea{\begin{eqnarray}}
\def\eea{\end{eqnarray}}	

\def\nn{\nonumber}

\newcommand{\GeV}{\;\text{GeV}}

\begin{document}
\begin{titlepage}
\vspace*{-1cm}
\phantom{hep-ph/***} 

\vskip 1.5cm
\begin{center}
{\LARGE\bf Higgs portal dark matter and neutrino mass and mixing with a doubly charged scalar}\\[3mm]
\vskip .3cm
\author{Last modified by: {\bf \USER }}
\end{center}
\vskip 0.5  cm
\begin{center}
{\large I.M.~Hierro}~$^{a)}$,
{\large S. F.~King}~$^{b)}$
{\large and S.~Rigolin}~$^{a)}$
\\
\vskip .7cm
{\footnotesize
$^{a)}$~
Dipartimento di Fisica ``G.~Galilei'', Universit\`a di Padova and \\
INFN, Sezione di Padova, Via Marzolo~8, I-35131 Padua, Italy \\
$^{b)}$~ Physics and Astronomy, University of Southampton,
\\
SO17 1BJ Southampton, United Kingdom \\
\vskip .1cm
\vskip .5cm
\begin{minipage}[l]{.9\textwidth}
\begin{center} 
\textit{E-mail:} 
\email{ignacio.hierro@pd.infn.it},
\email{s.f.king@soton.ac.uk},
\email{stefano.rigolin@pd.infn.it}.
\end{center}
\end{minipage}
}
\end{center}

\vskip 1cm
\abstract{ 
We consider an extension of the Standard Model involving two new scalar particles around the TeV scale: a singlet 
neutral scalar $\phi$, to be eventually identified as the Dark Matter candidate, plus a doubly charged $SU(2)_L$ 
singlet scalar, $S^{++}$, that can be the source for the non-vanishing neutrino masses and mixings. Assuming 
an unbroken $Z_2$ symmetry in the scalar sector, under which only the additional neutral scalar $\phi$ is odd, 
we write the most general (renormalizable) scalar potential. 
The model may be regarded as a possible extension of the conventional Higgs portal Dark Matter 
scenario which also accounts for neutrino mass and mixing.
This framework cannot completely explain the observed positron excess. However a softening of the 
discrepancy observed in conventional Higgs portal framework can be obtained, especially when the scale of new 
physics responsible for generating neutrino masses and lepton number violating processes is around 2 TeV.} 
\end{titlepage}
\setcounter{footnote}{0}


%
%
\newpage
\section{Introduction}

The present ensemble of data from accelerators experiment seems to firmly confirm all the Standard Model (SM) 
ingredients to high level of accuracy, including the presence of a relatively light scalar boson (so-called 
``Higgs'' for short) \cite{Aad:2012tfa,Chatrchyan:2012xdj}. 

There are, however, few experimental indications that there should be some type of new physics beyond the SM one. 
The most clear indication of the need of some new kind of matter, namely the Dark Matter (DM), derives from 
cosmological and astrophysical observations. Assuming the WIMP ansatz, the amount of measured DM density is 
consistent with the existence of a weakly interacting particle with a mass around the TeV scale.

On the other side, several theoretical features of the SM still need to be enlucidated, like for example the 
stability of the Higgs mass, the so called ``hierarchy problem'', or the presence of extremely different 
parameters describing the masses and mixings of the SM fermions, dubbed often as the ``flavour problem''. 



Following the idea that the fermion mass 
structures could arise from a symmetry principle, flavour symmetries have been introduced, both in the context of 
the SM and its extensions.
Many different examples have been proposed in the literature based 
on a large variety of symmetries: either abelian on non-abelian, local or global, continuous or discrete
\cite{King:2015aea}.
Despite of all these attempts, however, it seems unlikely that the same mechanism could be responsible to generate 
at the same time charged and neutral fermion masses. However 
the unique possibility of having Majorana masses for the 
neutrinos, associated with the exceedingly small values of the neutrino masses, could
be responsible for the differences observed in the 
neutrino flavour sector. 

The see-saw mechanism \cite{seesaw}, where heavy 
right-handed neutrinos with large Majorana masses are responsible for small effective left-handed neutrino masses, 
is notoriously difficult to directly test. Other mechanisms for generating neutrino masses include R-parity violating 
supersymmetry \cite{Hirsch:2000ef}, Higgs triplet models \cite{Magg:1980ut,Schechter:1980gr,Lazarides:1980nt}, or loop 
models involving additional Higgs doublets and singlets (e.g. \cite{Ma:2006km,Zee:1985id,Babu:1988ki}). 
All these models can be tested experimentally (for a review of these different mechanisms see for example 
\cite{Bandyopadhyay:2007kx}, and \cite{Law:2013dya} for a very systematic study). In particular, such settings 
can yield very interesting connections between lepton number violating physics and collider phenomenology 
\cite{Angel:2012ug,Babu:2001ex,Bonnet:2012kh}, especially if doubly charged scalars are involved (as in the 
Higgs triplet case  \cite{Perez:2008ha,Chun:2013vma}).

In this paper we shall focus on a particularly economical loop model of Majorana neutrino mass and mixing 
\cite{King:2014uha}, in which the low energy effective theory involves just one extra new particle: 
a doubly charged EW singlet scalar $S$ (denoting both $S^{++}$ and its antiparticle $S^{--}$). 
It is already known that such a model can lead to an interesting 
complementarity between low energy charged lepton flavour violation processes, and high energy collider physics,
depending on whether the doubly charged scalar $S$ appears as a virtual or real particle
\cite{Geib:2015tvt}. However such a model cannot account for DM, since the doubly charged scalar $S$ decays 
promptly into either pairs of like-sign charged leptons or $W$ bosons.
Here we shall extend the model slightly by introducing an additional neutral scalar $\phi$ and 
assume
an unbroken $Z_2$ symmetry in the scalar sector, under which only the additional neutral scalar $\phi$ is odd,
which then becomes a stable DM candidate. 
The model may be regarded as an extension of the so-called Higgs portal scenario \cite{Patt:2006fw},
in the presence of a doubly charged scalar which accounts for neutrino mass and mixing.
The resulting framework presented here, involving both $S$ and $\phi$, 
then merges two apparently unrelated features: the existence of a new physics 
sector at the TeV scale, providing naturally small neutrino masses, 
and the existence of a good DM candidate. 

The layout of the remainder of the paper is as follows.
In section~\ref{Stevemodel} we review the effective model 
proposed and studied in \cite{King:2014uha,Geib:2015tvt} involving just one extra particle,
the doubly charged scalar $S$. In section~\ref{Potential} we extend this model by introducing an additional
neutral scalar $\phi$, and discuss the resulting scalar potential of the model involving the Higgs doublet 
$H$, the doubly charged scalar $S$ and the neutral scalar $\phi$. We then go on to calculate the relic abundance 
of the DM particles $\phi$ and their prospects for direct and indirect detection.
Section~\ref{conclusions} concludes the paper.

%

\section{The effective model with a doubly charged scalar } 
\label{Stevemodel}
In this section we review the effective Lagrangian model presented in \cite{King:2014uha}, in which the SM is extended by adding 
one new scalar particle: a complex $SU(2)_L$ singlet, hypercharge $Y=2$ (hence electric charge $Q=2$) state 
$S^{++}$ and its antiparticle $S^{--}$, both doubly charged and denoted
collectively as $S$. 

The doubly charged scalar field $S$ has an effective coupling to the SM $W^{\pm}$ bosons as well as to same-sign right-handed charged 
SM leptons, giving rise to a rich phenomenology. In addition to contributing to flavour violating leptonic processes, to 
leptonic dipole moments and to leptonic radiative decays, the scalar $S$ allows a 2-loop diagram which 
is responsible for providing all mass (and mixings) to neutrinos. It 
is shown in \cite{King:2014uha} that the lowest mass dimension at which the vertex $SWW$ can be realised is by effective 
operators of dimension $d=7$. The relevant operator, in the unitary gauge, for the generation of neutrino masses is:
\bea
\LL_{SWW} &= & 
        - \dfrac{g^2\xi v^4}{4\Lambda^3}\left(S\,W^\mu W_\mu \,+\, h.c. \right)
\eea
being $\xi$ an order $\mathcal{O}(1)$ dimensionless parameter and $\Lambda$ the new physics scale above which the 
effective theory breaks. The coupling of $S$ to same-sign RH leptons is given by
\begin{equation}
\LL_{Sll}=f_{ab}\,S^\dagger\,\bar{l}_aP_Ll^c_b \,+\, h.c.
\end{equation}
with $f_{ab}$ dimensionless parameters. There are strong experimental constraints on the $f_{ab}$ parameter space, 
basically due to the flavour violating couplings of the charged scalar $S$ with leptons, the strongest bound 
proceeding from $\mu \rightarrow e \gamma$ and $\mu \rightarrow 3e$. A detailed analysis of these bounds can be 
found in \cite{King:2014uha,Herrero-Garcia:2014hfa,Geib:2015tvt}.

The simultaneous presence of the $SWW$ and $Sll$ vertices generate a 2-loop contribution to the neutrino masses, that 
schematically can be written as 
\begin{equation}
 \mathcal{M}_\nu^{\mathrm{2-loop}}=2\,\xi\,f_{ab}(1+\delta_{ab}) \dfrac{m_a\,m_b\,m^2_S }{\Lambda^3}\,
 \tilde{\mathcal{I}}(m_W,m_S,\mu) \,
\end{equation}
where $m_S$ is the S particle mass, $m_{i}$ is the mass of the $l_i$ lepton, 
$\delta_{ab}$ is the Kronecker delta and $\tilde{\mathcal{I}}(m_W,m_S,\mu)$ is 
the two loop integral calculated in \cite{King:2014uha}.

Apart from the usual contribution to $0\nu\beta\beta$ due to massive neutrinos in presence of a lepton number 
violating interaction, this model also produce an additional non-standard contribution to it, since the doubly 
charged scalar $S$ can couple both to $W^-W^-$ and $e^-e^-$. 
Taking into account the newest GERDA results of $T^{0\nu\beta\beta}_{1/2}\mathrm{(Ge)}>2.1\cdot 10^{25}$ at 
$90\%$ C.L., \cite{Agostini:2013mzu}, one obtains
\begin{equation}
\dfrac{\xi f_{ee}}{M^2_S\Lambda^3}<\dfrac{4.0\cdot 10^{-3}}{\mathrm{TeV}^5}\,.
\label{GERDA}
\end{equation}

In general it is not an easy task to fulfil all the flavour/dipole bounds and obtain a realistic description for 
neutrino masses and mixing compatible with the $0\nu\beta\beta$ decay bounds. In \cite{King:2014uha} a detailed 
analysis has been performed that highlighted the presence of three typical regions where this may happen, hereafter 
denoted as ``Benchmark Scenarios'':  
\begin{enumerate}
 \item Benchmark Scenario A: $f_{ee}\simeq0$ and $f_{e\tau}\simeq0$. In this region the additional contribution to 
       the $0\nu\beta\beta$ essentially vanishes. A normal hierarchy between the neutrino masses with the lightest 
       one around $5$ meV is obtained;
 \item Benchmark Scenario B: $f_{ee}\simeq 0$ and $f_{e\mu} \simeq -(f^*_{\mu\tau}/f^*_{\mu\mu}) f_{e\tau}$. In this 
       region one still has a vanishing additional contribution to the $0\nu\beta\beta$ and a normal ordered neutrino 
       masses with the lightest one around $5$ meV. However the constraint relating $f_{e\mu}$ and $f_{e\tau}$ makes 
       this scenario more predictive (falsifiable) in what concerns lepton flavour violation;  
 \item Benchmark Scenario C: $f_{e\mu}\simeq -(f^*_{\mu\tau}/f^*_{\mu\mu}) f_{e\tau}$. In this region one can assume 
       large values for the $f_{ee}$ coupling. However not to enter in conflict with the GERDA limit on $0\nu\beta\beta$ 
       of Eq.~(\ref{GERDA}) one has to push the cutoff scale $\Lambda$ to several TeV, not a desirable thing from the 
       collider phenomenology point of view.
\end{enumerate}

For the analysis presented in the following sections we will use the best fit benchmark point for each of the three 
scenarios reported by \cite{King:2014uha}:
\begin{enumerate}
 \item Benchmark Point A: $m_S=164.5$ GeV, $\Lambda=905.9$ GeV , $\xi=5.02$;
 \item Benchmark Point B: $m_S=364.6$ GeV, $\Lambda=2505.1$ GeV, $\xi=6.38$;
 \item Benchmark Point C: $m_S=626.0$ GeV, $\Lambda=5094.7$ GeV, $\xi=3.39$.
\end{enumerate}

%
%
\section{Higgs portal DM with a doubly charged scalar} 
\label{Potential}

In order to account for DM, we now introduce a further particle into the scheme of the previous section, 
namely an electrically neutral real scalar $\phi$.
An unbroken $\mathbb{Z}_2$ 
symmetry is assumed, under which the field $\phi$ is odd, while all the other particles are even. The motivation of 
such a setup is twofold: firstly, as already discussed, the presence of an extra doubly charged scalar can provide an economical mechanism 
for triggering light neutrino masses and mixing \cite{King:2014uha,Geib:2015tvt,Babu:2002uu,Gustafsson:2012vj}. 
Secondly, the new neutral scalar can account for DM. Possible UV completions 
of this model could be pursued along the lines of \cite{Babu:2002uu,Gustafsson:2012vj}. 
Here we shall not try to construct an ultraviolet complete model, but continue to consider the effective theory
below the cut-off $\Lambda$, where the theory has a rather minimal particle content,
with the goal of understanding DM in this extended model.
In particular, we shall discuss how 
the presence of an extended scalar sector can potentially modify the limits and the predictions obtained in 
the standard DM Higgs portal scenario \cite{Patt:2006fw}. In this section we study the DM signatures of the effective Lagrangian 
model described in the previous section.

The most general (renormalizable) scalar potential for the model at hand is given by
\bea
V &=& \mu^2|H^\dagger H| + \lambda|H^\dagger H|^2 + \dfrac{1}{2}\mu^2_\phi\phi^2 + \dfrac{1}{4}\lambda_\phi\phi^4 
      + \mu^2_S S^\dagger S  + \lambda_S (S^\dagger S)^2+ \nn \\
  & & + \dfrac{1}{2}\lambda_{\phi H}\phi^2\,|H^\dagger H| + \lambda_{S H}(S^\dagger S)\,|H^\dagger H| +
      \dfrac{1}{2}\lambda_{\phi S}\phi^2\,(S^\dagger S)
\label{scalarpot}
\eea
where $H$ is the usual SM Higgs doublet, $S$ the doubly charged scalar and $\phi$ the additional neutral scalar, odd 
under the unbroken $Z_2$ symmetry, that will play eventually the role of stable DM. In addition to the SM Higgs 
sector parameters, $\mu$ and $\lambda$, compatibly with the assumption of an unbroken $\mathbb{Z}_2$ symmetry, one 
can introduce seven additional dimensionless parameters: a quadratic and a quartic self--interacting couplings, 
respectively for the neutral and charged exotic scalars, plus three parameters associated to the quartic mixings 
between all the neutral and charged scalars. We assume that the ElectroWeak Symmetry Breaking (EWSB) is associated 
exclusively to the Higgs sector, i.e. $\mu^2 < 0$ is assumed while $\mu_\phi^2,~\mu_S^2 > 0$ are considered. The 
masses of the exotic scalars, then, read
\bea  
m^2_\phi \equiv \mu^2_\phi+\dfrac{1}{2}\lambda_{\phi H}v^2 \qquad , \qquad 
m^2_S \equiv \mu^2_S+\dfrac{1}{2}\lambda_{S H}v^2
\label{exoticmasses}
\eea

It is interesting to compare the predictions of this model with the ones of the minimal Higgs portal DM, 
which is described by the potential of Eq.~(\ref{scalarpot}) once the doubly charged scalar is decoupled 
from the theory, i.e. $m_S \gg m_H, m_\phi$ or by setting $\lambda_{\phi S}=0=\lambda_{S H}$\footnote{Notice, 
however, that the decoupling limit is only approximately reached by setting one of the tree level parameters, 
for example $\lambda_{\phi S}$, to zero, while keeping the other two finite. In fact in this case one can generate 
a one--loop contribution to $\lambda_{\phi S}$ through the $\lambda_{\phi H}$ and $\lambda_{S H}$ vertices. 
We will come back later on this point.}. 
The phenomenology of such a minimal Higgs portal DM model has been extensively studied in \cite{Silveira:1985rk,McDonald:1993ex,
Burgess:2000yq,Davoudiasl:2004be,Ham:2004cf,Patt:2006fw,O'Connell:2006wi,He:2007tt,Profumo:2007wc,Barger:2007im,
He:2008qm,Ponton:2008zv,Lerner:2009xg,Farina:2009ez,Bandyopadhyay:2010cc,Barger:2010mc,Guo:2010hq,Espinosa:2011ax,
Profumo:2010kp,Djouadi:2012zc,Mambrini:2012ue,Drozd:2011aa,Grzadkowski:2009mj,Cline:2013gha}. 
Here we are interested in how the presence of the doubly charged scalar can affect Higgs portal DM.


%
\subsection{Relic abundance}

In order to obtain the DM relic abundance one has to solve the following Boltzman equation:
\begin{equation}
\dfrac{dY}{dT}=\sqrt{\dfrac{\pi g_*(T)}{45}}M_P\braket{\sigma v}\left(Y(T)^2-Y_{eq}(T)^2\right) 
\label{boltz}
\end{equation}
where $Y(T)$ is the DM abundance, $Y_{eq}(T)$ is the equilibrium thermal abundance, $g_*$ is the effective number 
of degrees of freedom, $M_P$ is the Planck mass and $\braket{\sigma v}$ is the thermally averaged annihilation 
cross section, which must include all relevant annihilation processes:
\begin{equation}
\braket{\sigma v}=\int_{4m^2_\phi}^\infty \dfrac{s\sqrt{s-4m^2_\phi}K_1(\sqrt{s}/T)\sigma v_{rel}}
                  {16Tm^4_\phi K^2_2(m_\phi/T)}ds
\label{thermalaverage}                  
\end{equation}
where $K_1$ and $K_2$ are modified Bessel functions of the second kind. The present DM abundance, $Y(T_0)$, 
is obtained by integrating Eq.~(\ref{boltz}) down to the today temperature $T_0$. Then, the DM relic density is
\begin{equation}
\Omega_{DM} h^2=2.74\times10^8\dfrac{m_\phi}{\mathrm{GeV}}Y(T_0) \,.
\end{equation}
We computed these quantities using the publicly available version of micrOMEGAs \cite{Belanger:2004yn,Belanger:2006is}.


\begin{figure}[t!]
\centering{
\begin{tabular}{ c }
\hspace{-4ex}\includegraphics[scale=1.4]{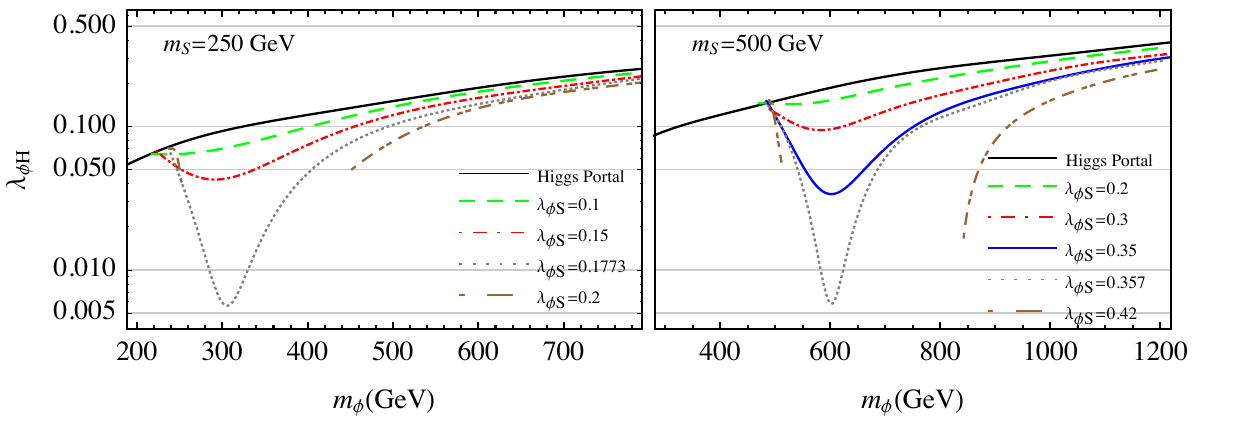} \\
\hspace{-4ex}\includegraphics[scale=1.4]{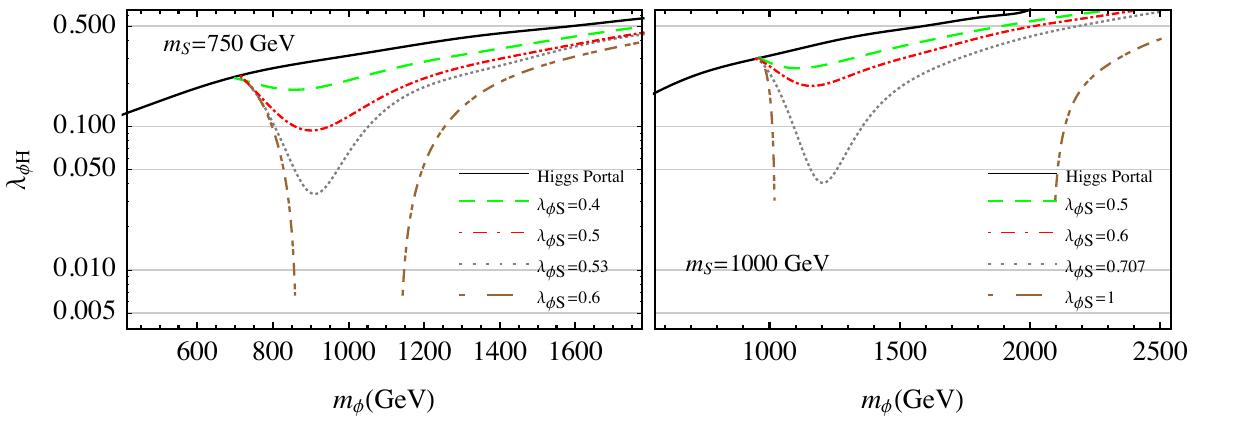} \\
\end{tabular}}
\caption{
Allowed parameter space, in the ($m_\phi$,$\lambda_{\phi H}$) plane, for which the DM relic abundance reproduces 
the observed value, $\Omega_{DM} h^2=0.1198$ \cite{Ade:2015xua}. The four plots correspond to four different values 
of $m_S=250,500,750,1000$ GeV, respectively. In each plot several choices for $\lambda_{\phi S}$ are shown. }
\label{lahphmph}
\end{figure}


In Fig.~\ref{lahphmph} we plot the allowed parameter space in the ($m_\phi$,$\lambda_{\phi H}$) plane for which 
the DM relic abundance coincides with the observed value, $\Omega_{DM} h^2=0.1198$ \cite{Ade:2015xua}. The four plots 
correspond to four different values of $m_S=250,500,750,1000$ GeV, respectively. In each plot of Fig.~\ref{lahphmph}, 
the full black curve represents the Higgs portal case (i.e. $\lambda_{\phi S}=0=\lambda_{S H}$). Then, for each plot 
the dashed, dot--dashed, dotted and dot--dot--dashed curve are obtained for representative choices of $\lambda_{\phi S}$, 
which value is shown in the legenda. There is no significative dependence from the choosen value of $\lambda_{S H}$, 
which has been conventionally taken $\lambda_{S H}=1$.\footnote{On the one hand, since the $\phi\phi$ annihilation 
occurs almost at rest, $m_S>m_H$ in all the considered scenarios, the $H\rightarrow S S^{\dagger}$ decay is not relevant. 
On the other hand, the contributions coming from $S S^{\dagger}\rightarrow H $ are suppressed by a loop factor.}

When the DM particle, $\phi$, is lighter than the doubly charged scalar, $S$, the process $\phi\phi\rightarrow 
S S^{\dagger}$ is not efficient, and the relic abundance results as in the pure Higgs portal case, i.e. via the 
$\phi\phi\rightarrow H$ annihilation process. The same happens when $m_\phi\gg m_S$. However, in the intermediate 
region, $m_\phi \approx m_S$, the process $\phi\phi\rightarrow S S^\dagger$ becomes efficient and, accordingly, 
$\lambda_{\phi H}$ needs to be suppressed, depending on the chosen value for $\lambda_{\phi S}$, in order to reproduce 
the correct amount of DM relic density. In particular, for large enough $\lambda_{\phi S}$ and for specific values of 
the $m_\phi$ mass all the relic abundance can be produced exclusively via the coupling $\lambda_{\phi S}$, with 
$\lambda_{\phi H}$ approaching zero. This is why the dot--dot--dashed (brown) curve exists only 
in the ``small'' and ``large'' $m_\phi$ region. For such values of $\lambda_{\phi S}$, in the intermediate $m_\phi$ range, 
DM is overproduced, and the set of parameter chosen is not allowed. This fact can be clearly seen in the $m_S=750$ GeV 
(lower-left) plot: for $\lambda_{\phi S} = 0.6$ and $850 \lesssim m_\phi \lesssim 1150$ GeV one can never reproduce 
the correct amount of DM density.


A comment is in order regarding the possibility of setting $\lambda_{\phi H}$=0. 
In the framework at hand, $\lambda_{\phi H}$ can receive loop contributions, through diagrams involving $\lambda_{\phi S}$ 
and $\lambda_{H S}$ couplings. It can be shown that, for a temperature $T>m_\phi/20$, $\braket{\sigma v}$ (and 
therefore $\lambda_{\phi H}$) doesn't have an impact on $Y(T)$, since for those temperatures the DM abundance is 
equal to the equilibrium abundance, $Y(T)=Y_{eq}(T)$. When $T\sim m_\phi/20$, the DM particle freezes-out, and the 
relic abundance depends indeed on $\braket{\sigma v}$. As a conclusion, the typical energies in which the loop is 
relevant is when $p^2\sim (m_\phi /20)^2$. Setting the renormalization scale at $2m_S$, at one loop one obtains: 
\begin{equation}
\lambda_{\phi H}^{ren} \simeq \lambda_{\phi H}+\dfrac{1}{{16\pi^2}}\lambda_{\phi S}\lambda_{S H}\log\dfrac{m_\phi}{ 40m_S}
\end{equation}
As typically the loop contribution is few $10^{-3}$, one cannot extrapolate the tree level analysis to values of 
$\lambda_{\phi H}$ below few $10^{-3}$. In plotting our results we always work with $\lambda_{\phi H} \ge 0.005$. 

\begin{figure}[t!]
\centering
\includegraphics[scale=1.2]{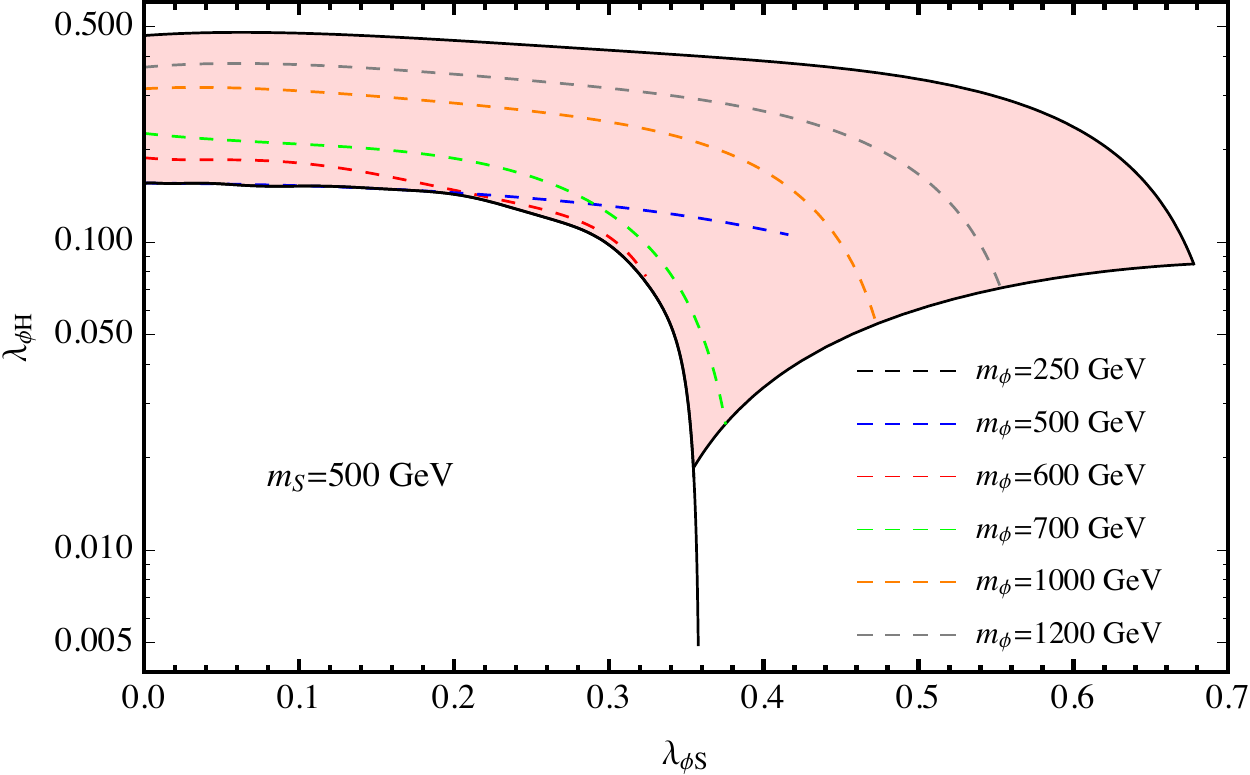}
\caption{Allowed parameter space in the ($\lambda_{\phi S}$,$\lambda_{H\phi}$) plane for $m_S=500\ \mathrm{GeV}$
and $m_\phi\in(500,1500)\mathrm{GeV}$. The value for $m_\phi=250 \ \mathrm{GeV}$ is shown in order to illustrate the 
fact that, for $m_\phi$ below $m_S$, the coupling $\lambda_{\phi S}$ has no impact.\protect}
\label{lasphlahph}
\end{figure}

All these comments are clearly summarised in Fig.~\ref{lasphlahph} where the parameter space which yields the correct 
relic abundance, in the ($\lambda_{\phi S}$,$\lambda_{H\phi}$) plane, is shown for $m_S=500$ GeV. The light-red region 
summarises the region, allowed by relic density data, for the relevant couplings of our DM model. Inside the filled 
region for definiteness we have also shown few dashed lines for various $m_\phi$ values. For $m_\phi \le m_S$ one 
typically spans the lower--left region of the parameter space, while for $m_\phi \ge m_S$ one spans the upper and 
the right part of the filled area. In particular, we clearly see the existence of a critical value: 
$\lambda^{crit}_{\phi S}=0.357$ for this specific $m_S$ case. For $\lambda_{\phi S} \le \lambda^{crit}_{\phi S}$ it 
is always possible to find values for $\lambda_{\phi H}$ and $\lambda_{\phi S}$ in order to satisfy the relic 
abundance bound, independently of the $m_\phi$ mass. In fact one always cross all different colours dashed lines. 
For $\lambda_{\phi S} \ge \lambda^{crit}_{\phi S}$, only for specific ranges of $m_\phi$ one can find a solution. 

\subsection{Direct detection}

\begin{figure}[ht!]
\includegraphics[scale=1.25]{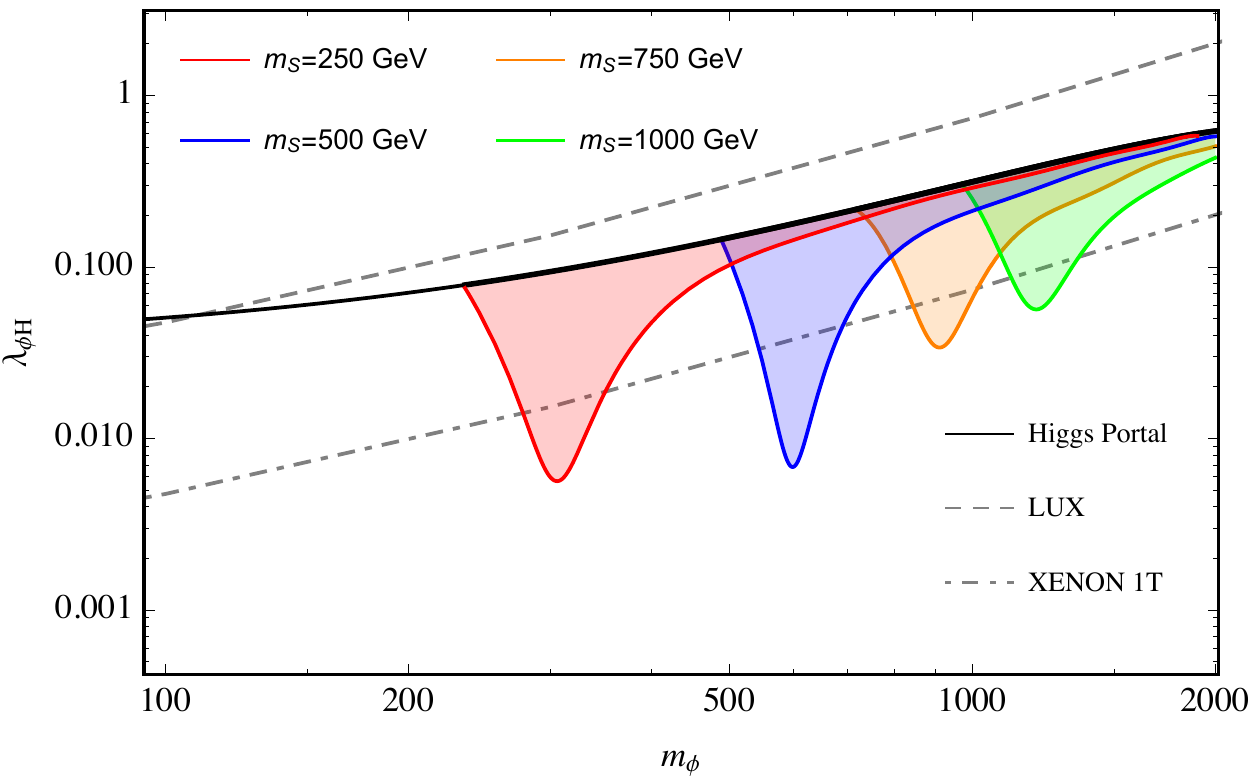}
\hspace{-0.5cm}
\caption{Allowed parameter space in the ($\lambda_{H\phi}$, $m_\phi$) plane for four different $m_S$ values ($m_S=250,
500,750,1000$ GeV). Dashed and dot--dashed black lines represent the exclusion limits from direct detection by LUX and 
(the prediction for) XENON 1T, respectively.\protect}
\label{direct}
\end{figure}

Direct detection experiments can significantly constrain the allowed parameter space for DM models. Experiments 
like LUX \cite{Akerib:2013tjd} and XENON \cite{Aprile:2012nq,Aprile:2012zx} can detect the DM particle scattering  
with the nucleons of the detector material, which in the both cases is Xenon. In our model, as well as in the pure 
Higgs portal case, this interaction mainly occurs via exchange of a Higgs scalar. The spin-independent cross-section 
is given by
\begin{equation}
\sigma_{SI}=\dfrac{f_N^2\mu^2m_N^2}{4\pi m^4_H m_\phi^2}\, \lambda_{\phi H}^2
\end{equation}
with $m_N$ the nucleon mass, $\mu=m_\phi m_N/(m_\phi+m_N)$ the DM-nucleon reduced mass and $f_N\sim0.3$ the hadron 
matrix element \cite{Cline:2013gha}. In Fig.~\ref{direct} we show the limits in the plane ($\lambda_{\phi H}$, $m_\phi$) 
from the current constraints of LUX (dashed black line) and the predicted sensitivity of XENON 1T (dot dashed black line). 
Both the pure Higgs portal and our model can escape the LUX limit. However, while the Higgs portal scenario will be for 
sure inside the XENON 1T sensitivity region, our model can for all considered values of $m_S$, ranging from $250$ GeV 
and $1000$ GeV, escape the direct detection. As explicitly shown in Figs.~\ref{lahphmph} and \ref{lasphlahph}, the 
presence of the new coupling $\lambda_{\phi S}$ can allow values for $\lambda_{\phi H}$ below XENON 1T sensitivity. 
This feature is almost independent from the chosen $m_\phi$ and $m_S$ values, in the $\approx$ 1 TeV region.

%
\subsection{Indirect detection}

The DM particles in the galactic centre can annihilate and yield several indirect signatures, such as positrons, 
antiprotons and photons. The detection of these cosmic rays is one of the most promising ways to identify DM existence 
\cite{Salati:2010rc,Porter:2011nv,Yaguna:2008hd,Goudelis:2009zz,Arina:2010rb}. We will focus here in particular in 
the observed spectrum of positrons and antiprotons. The production rate of these particles at a position $\vec{x}$ 
with an energy $E$ is usually expressed as \cite{Belanger:2010gh}
\begin{equation}
Q_a(\vec{x},E)=\dfrac{1}{2}\braket{\sigma v}\left(\dfrac{\rho(\vec{x})}{m_\phi} \right)^2 f_a(E)
\end{equation}
where $\sigma v$ is defined in Eq.~(\ref{thermalaverage}), $\rho(\vec{x})$ is the DM density and $f_a(E)=dN_a/dE$ is 
the energy distribution of the species $a$ produced in a single annihilation event. 

The region of diffusion of cosmic rays is represented by a disk of thickness $2 L \simeq (2-30)$ kpc and radius 
$R\simeq20$ kpc. The galactic disk is modelled as an infinitely thin disk lying in the middle with half-width $h=100$ pc 
and radius $R$. The charged particles, generated from DM annihilation, propagate in a turbulent regime through the strong galactic 
magnetic field and are deflected by its irregularities. Monte Carlo simulations show that this motion can be described 
by an energy dependent diffusion term $K(E)$. On top of that, these particles can lose their energy via inverse Compton 
scattering on interstellar medium, through Coulomb scattering or adiabatically. This energy loss rate is denoted by $b(E)$. 
Furthermore these particles can be wiped away by galactic convection, with a velocity $V_C\simeq(5-15)$ km/s 
\cite{Goudelis:2009zz}. Finally, one has also to account for the annihilation rate $\Gamma_{ann}$ induced by the 
interaction of the charged particles with ordinary matter in the galactic disk. Taking into account all these effects, 
the equation that describes the evolution of the energy distribution of charged particles reads:
\begin{equation}
\dfrac{\partial}{\partial z}\left(V_C\psi_a\right)-\mathrm{\nabla}\cdot\left(K(E)\mathrm{\nabla}\psi_a\right)-
\dfrac{\partial}{\partial E}\left(b(E)\psi_a\right)- 2h\delta(z)\Gamma_{ann}\psi_a=Q_a(\vec{x},E)
\end{equation}
where $z$ is the height in cylindrical coordinates adapted to the disk diffusion model, $\psi_a=dn/dE$ is the number 
density of particles per unit volume and energy. We use the default settings of micrOMEGAS \cite{Belanger:2010gh} 
to numerically evaluate the propagation of positrons and antiprotons that originate from DM annihilation.

\subsubsection{Positrons}

The energy spectrum of positrons originated from DM annihilation in the galactic centre is obtained by solving the 
diffusion-loss equation keeping only the two dominant contributions: space diffusion and energy losses,
\begin{equation}
-\mathrm{\nabla}\cdot\left(K(E)\mathrm{\nabla}\psi_{e^+}\right)-\dfrac{\partial}{\partial E}\left(b(E)\psi_{e^+}\right)
 = Q_{e^+}(\mathrm{x},E)
\end{equation}
The flux of $e^+$ originated from DM is then given by 
\begin{equation}\Phi_{e^+}^{DM}(E)=\dfrac{c}{4\pi}\psi_{e^+}(E,r_{\astrosun})
\end{equation}
where $c$ is the speed of light and $r_{\astrosun}\simeq8.5 \mathrm{kpc}$ is the distance from the Milky Way centre 
to the Sun.

In addition to the $e^+$ flux from the DM decay, there exists a secondary positron flux from interactions 
between cosmic rays and nuclei in the interstellar medium. This positron background $\Phi^{bg}_{e^+}$ can be well 
approximated as \cite{Strong:2004de,Baltz:1998xv}
\begin{equation}
\Phi^{bg}_{e^+}(E)=\dfrac{4.5 \cdot 10^{-4} \,E^{0.7}}{1+650E^{2.3}+1500E^{4.2}}[\mathrm{GeV^{-1} m^{-2}s^{-1}sr^{-1}}]
\label{positronback}
\end{equation}

\begin{figure}[t!]
\hspace{-0.9cm}
\includegraphics[scale=1.4]{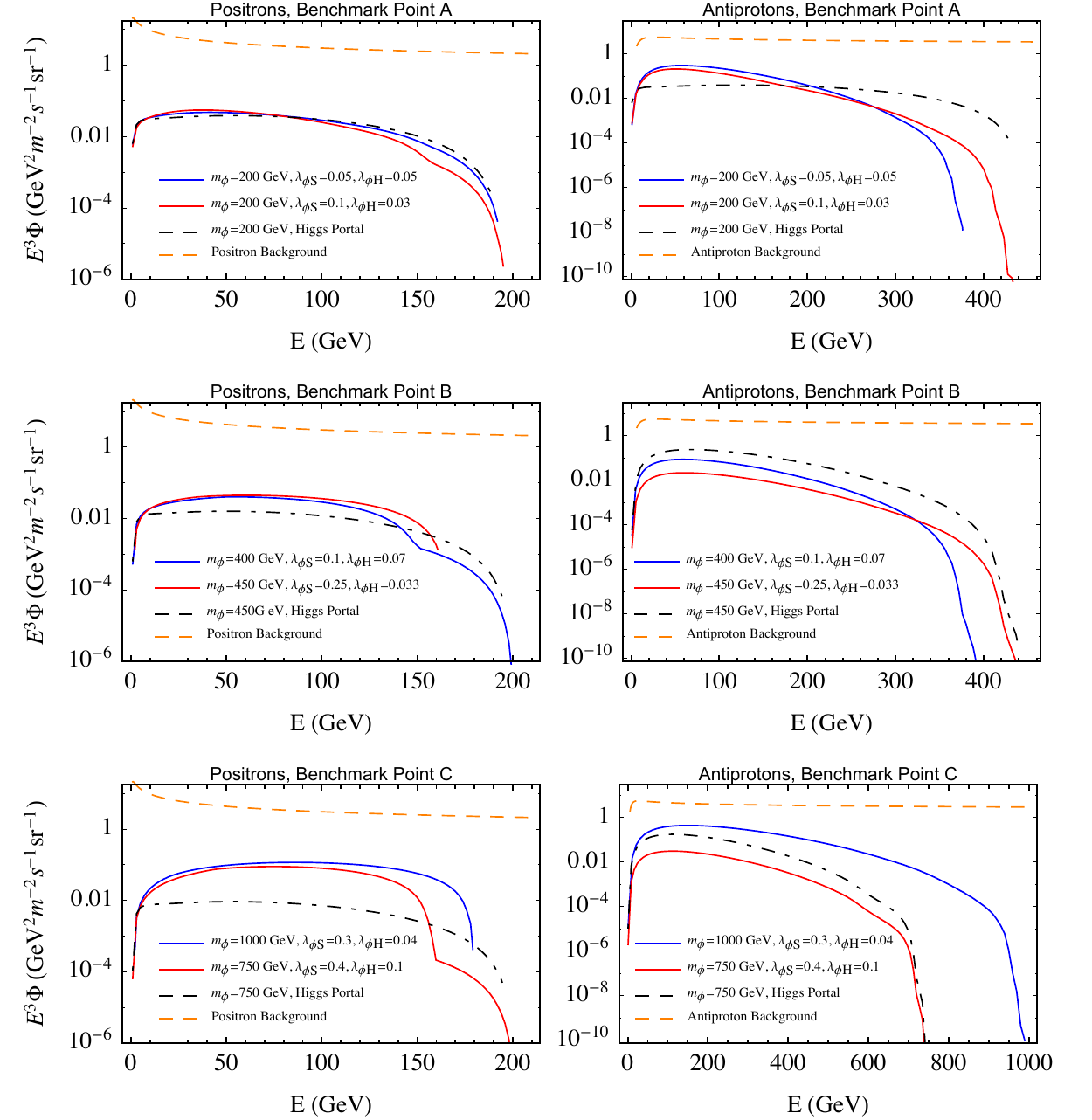}
\caption{Predicted positron (left column plots) and antiproton (right column plots) fluxes for the chosen benchmark 
points A,B and C respectively. \protect}
\label{allindirect}
\end{figure}

In the left column of Fig.~\ref{allindirect} we show the positron flux as function of the positron energy, for the three 
benchmark points mentioned in Section \ref{Stevemodel}. In each of the three left plots, the dashed (orange) line represents 
the expected background of Eq.~(\ref{positronback}), while the dot--dashed (black) line is the prediction for the Higgs 
portal case. The (blue and red) continuous lines represent our model expectations for two different choices of parameters, 
reported in each plot legenda, which give the correct relic abundance. 

The Higgs portal prediction is always at least two orders of magnitude below the astrophysical positron background. For the 
set of parameters defining Benchmark Point A, the expected positron flux almost coincides with the Higgs portal scenario one. 
This is due to the fact that for such values of $m_\phi$ and $m_S$, the $\phi\phi\rightarrow S S^\dagger$ channel is still 
suppressed compared to the usual $\phi\phi\rightarrow H$ one. Morever, for this Benchmark Point the S coupling to electrons 
and positrons $f_{ee} \approx 0$. For Benchmark Point B (middle left plot) the $\phi\phi\rightarrow S S^\dagger$ channel 
becomes more effective compared to the $\phi\phi\rightarrow H$ one, suppressed by the large $m_\phi$ mass. Still one has 
$f_{ee} \approx 0$, the dominant $S$ decays being in $WW$ and $e\tau$ (see \cite{King:2014uha}). This result in a positron 
flux two or three times higher than in the Higgs Portal scenario. Finally, for Benchmark Point C (lower left plot) the 
$\phi\phi\rightarrow S S^\dagger$ channel becomes dominant. Moreover, in this case one has sizeable $f_{ee}$, letting $S$ 
mostly decays in positrons. In this region of the parameter space, our model positron flux is one order if magnitude higher 
compared with the standard Higgs Portal scenario, even if still one order below the expected background. 

\subsubsection{Antiprotons}

The propagation of antiprotons originated from DM annihilation, neglecting the energy loss term, can be described as 
\cite{Cao:2014cda}
\begin{equation}
-K_{\bar{p}}\nabla^2\psi_{\bar{p}}+V_C\dfrac{\partial}{\partial z}\psi_{\bar{p}}+
   2h\delta(z)\Gamma_{ann}\psi_{\bar{p}}=Q_{\bar{p}}(\mathrm{x},E)
\end{equation}
The astrophysical antiproton background $\Phi_{\bar{p}}^{bg}$ can be written as 
\begin{equation}
\Phi^{bg}_{\bar{p}}=\dfrac{0.9E^{-0.9}}{14+30E^{-1.85}+0.08E^{2.3}}[\mathrm{GeV^{-1}m^{-2}s^{-1}sr^{-1}}]
\label{antiprotonback}
\end{equation}

The plots on the right column of Fig.~\ref{allindirect} show that, as in the positron case, the antiproton flux 
predicted in the Higgs portal scenario (dot--dashed black line) is roughly two orders of magnitude below the 
astrophysical background (dashed orange curve). 

The doubly charged particle has a largest branching fraction to $W$s in Benchmark Point A than in the rest of 
cases \cite{King:2014uha}. Thus, even if in this region of parameter space the $\phi\phi\rightarrow H$ process 
still dominates, the flux of antiprotons for $E \le 200$ GeV is higher than the one predicted in the Higgs Portal 
case. However, for Benchamrk Points B and C, where the $S$ scalar decays mostly to leptons ($e\tau$ and $ee$, 
respectively), the increasing relevance of the $\phi\phi\rightarrow S S^\dagger$ process makes the $\bar{p}$ flux 
smaller than the corresponding Higgs Portal case for the same $m_\phi$ mass (compare the black and red lines in 
middle and bottom left plots of Fig.~\ref{allindirect}, respectively for $m_\phi=450\GeV$ and $m_\phi=750\GeV$).

In our model one can obtain a larger flux by increasing the $\phi$ mass. For example for Benchmark Point C (bottom 
left plot in  Fig.~\ref{allindirect}), one can obtain a rather larger contribution to the antiproton flux selecting 
$m_\phi =1000$ GeV, but still one order of magnitude smaller than the expected $\bar{p}$ background.

\section{Conclusions}
\label{conclusions}

In this letter we have considered an extension of the Standard Model involving two new scalar particles around the 
TeV scale: a singlet neutral scalar $\phi$, that plays the role of the Dark Matter candidate plus a doubly charged 
$SU(2)_L$ singlet scalar, $S^{++}$, that is the source for the non-vanishing neutrino masses and mixings. In this 
framework, besides being able to explain naturally the smallness of neutrino masses with the new physics at the TeV 
scale, it could be possible to identify DM scenarios which extend the conventional Higgs portal one. We have 
studied the allowed parameter space for our model, compatible with the present DM relic density. Moreover we have 
identified possible signatures from direct and indirect DM detection experiments. In general our results indicate that it would is possible, within our framework, 
to evade XENON 1T exclusion limits for a significant region of the parameter space. However, we also show that,
even if the positron and antiproton flux, originating from DM annihilation in the centre of our galaxy, 
is higher than the standard Higgs portal one, it is still about an order of magnitude lower then the observed background.  

In conclusion, our model may be regarded as an extension of the minimal Higgs portal DM scenario with a doubly charged scalar
which can account for neutrino mass and mixing. The presence of the doubly charged scalar $S$ introduces a new
portal coupling of the DM particle $\phi$ to $S$, namely $\lambda_{\phi S}$, in addition to the usual Higgs portal
coupling of $\phi$ to the Higgs doublet $H$, $\lambda_{H\phi}$. The new portal coupling $\lambda_{\phi S}$
becomes important when $m_\phi$ exceeds $m_S$, since then it allows the DM particle to annihilate into pairs
of doubly charged scalars, as an alternative to the usual DM annihilation into Higgs pairs. This in turn reduces the 
coupling $\lambda_{H\phi}$, consistent with the desired relic density, making DM harder to detect by direct detection experiments.

\section*{Acknowledgements}

The authors acknowledge support from the
European Union Horizon 2020 research and innovation programme under the Marie 
Sklodowska-Curie grant agreements   InvisiblesPlus RISE No. 690575, Elusives ITN No. 674896 and 
Invisibles ITN No. 289442.
SFK acknowledge support from the STFC Consolidated grant ST/L000296/1.



\end{document}